\pdfoutput=1
\documentclass{article}

\usepackage[final]{neurips_2023}

\usepackage[utf8]{inputenc} 
\usepackage[T1]{fontenc}    
\usepackage{hyperref}       
\usepackage{url}            
\usepackage{booktabs}       
\usepackage{amsfonts}       
\usepackage{nicefrac}       
\usepackage{microtype}      
\usepackage{xcolor}         
\usepackage{xspace}

\usepackage{tikz,pgfplots}
\pgfplotsset{compat=newest}
\pgfplotsset{/pgfplots/error bars/error bar style={very thick}}

\usepackage{multirow}
\usepackage{graphicx}
\usepackage{caption}
\usepackage{subcaption}
\usepackage{pifont}
\usepackage{wrapfig}
\usepackage{amsmath, amssymb, amsthm}
\newcommand{\pmr}[1]{\scriptsize$\pm$#1}
\newcommand{\method}{\textsc{MusicGen}\xspace}
\usepackage[capitalise]{cleveref}
\newcommand{\encodec}{EnCodec }
\newcommand{\real}{\mathbb{R}}
\newcommand{\proba}[1]{\mathbb{P}\left[#1\right]}
\newcommand{\newparagraph}[1]{\noindent {\bf #1}}
\newcommand{\chmark}{\ding{51}}
\newcommand{\crmark}{\ding{55}}
\usepackage{etoolbox}
\newtoggle{release}
\togglefalse{release}
\toggletrue{release}

\iftoggle{release}{
\newcommand{\alex}[1]{}
\newcommand{\gab}[1]{}
\newcommand{\adios}[1]{}
\newcommand{\jade}[1]{}
}
{
\newcommand{\alex}[1]{{\color{blue} A: #1}}
\newcommand{\gab}[1]{{\color{red} G: #1}}
\newcommand{\adios}[1]{{\color{magenta}Y: #1}}
\newcommand{\jade}[1]{{\color{cyan}J: #1}}

}

\hypersetup{
    colorlinks,
    linkcolor={red!50!black},
    citecolor={blue!50!black},
    urlcolor={blue!80!black}
}

\Crefname{figure}{Figure}{Figures}

\title{Simple and Controllable Music Generation}

\author{
\quad\textbf{Jade Copet}$^{\spadesuit\diamondsuit}$ 
\quad\textbf{Felix Kreuk}$^{\spadesuit\diamondsuit}$
\quad\textbf{Itai Gat}
\quad\textbf{Tal Remez}
\quad\textbf{David Kant}\\
\quad\textbf{Gabriel Synnaeve} $^\diamondsuit$
\quad\textbf{Yossi Adi\thanks{Yossi Adi is Affiliated with both The Hebrew University of Jerusalem \& MetaAI.}} $^\diamondsuit$
\quad\textbf{Alexandre D\'efossez} $^\diamondsuit$ \\ 
{\footnotesize $\spadesuit$: equal contributions, $\diamondsuit$: core team}\\
\vspace{0.05cm}
Meta AI\\
{\tt \{jadecopet, felixkreuk, adiyoss\}@meta.com}}

\begin{document}

\maketitle

\begin{abstract}
    We tackle the task of conditional music generation. We introduce \method, a single Language Model (LM) that operates over several streams of compressed discrete music representation, i.e., tokens. Unlike prior work, \method is comprised of a single-stage transformer LM together with efficient token interleaving patterns, which eliminates the need for cascading several models, e.g., hierarchically or upsampling. Following this approach, we demonstrate how \method can generate high-quality samples, both mono and stereo, while being conditioned on textual description or melodic features, allowing better controls over the generated output. We conduct extensive empirical evaluation, considering both automatic and human studies, showing the proposed approach is superior to the evaluated baselines on a standard text-to-music benchmark. Through ablation studies, we shed light over the importance of each of the components comprising \method. Music samples, code, and models are available at \href{https://github.com/facebookresearch/audiocraft}{github.com/facebookresearch/audiocraft}.
\end{abstract}

\section{Introduction}\label{sec:intro}

Text-to-music is the task of generating musical pieces given text descriptions, e.g., ``90s rock song with a guitar riff''. Generating music is a challenging task as it requires modeling long range sequences. Unlike speech, music requires the use of the full frequency spectrum~\citep{muller2015fundamentals}. That means sampling the signal at a higher rate, i.e., the standard sampling rates of music recordings are 44.1 kHz or 48 kHz vs. 16 kHz for speech. Moreover, music contains harmonies and melodies from different instruments, which create complex structures. Human listeners are highly sensitive to disharmony~\citep{fedorenko2012sensitivity, norman2019divergence}, hence generating music does not leave a lot of room for making melodic errors. Lastly, the ability to control the generation process in a diverse set of methods, e.g., key, instruments, melody, genre, etc. is essential for music creators. 

Recent advances in self-supervised audio representation learning~\citep{balestriero2023cookbook}, sequential modeling \citep{touvron2023llama}, and audio synthesis~\citep{tan2021survey} provide the conditions to develop such models. To make audio modeling more tractable, recent studies proposed representing audio signals as multiple streams of discrete tokens representing the same signal~\citep{defossez2022highfi}. This allows both high-quality audio generation and effective audio modeling. However, this comes at the cost of jointly modeling several parallel dependent streams.

\citet{kharitonov2022text, kreuk2022audiogen} proposed modeling multi-streams of speech tokens in parallel following a delay approach, i.e., introduce offsets between the different streams. \citet{agostinelli2023musiclm} proposed representing musical segments using multiple sequences of discrete tokens at different granularity and model them using a hierarchy of autoregressive models. In parallel, \citet{donahue2023singsong} follows a similar approach but for the task of singing to accompaniment generation.  Recently, \citet{wang2023neural} proposed tackling this problem in two stages: (i) modeling the first stream of tokens only; (ii) then, apply a post-network to jointly model the rest of the streams in a non-autoregressive manner.

\begin{figure*}[t!]
    \centering
    \includegraphics[width=0.9\textwidth]{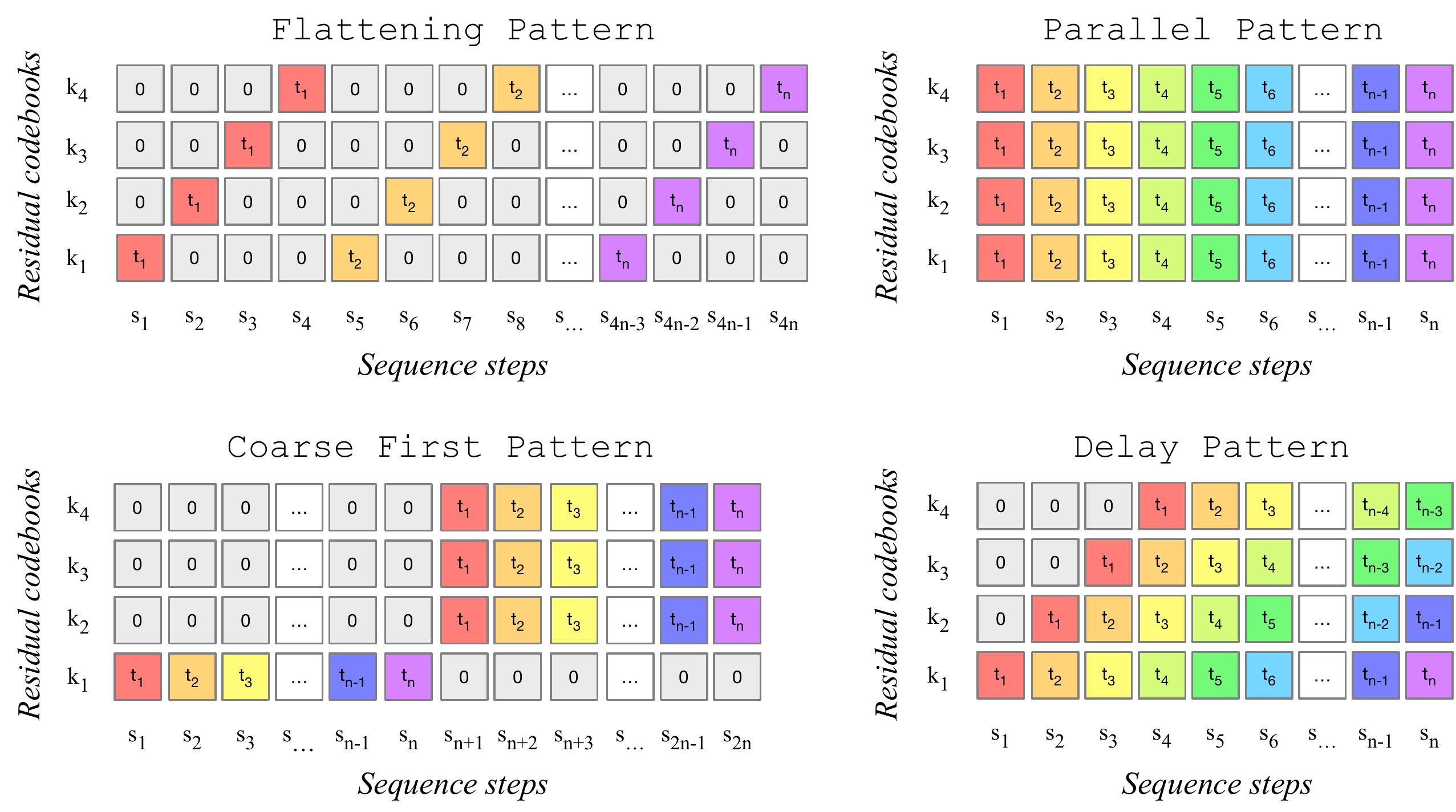}
    \caption{Codebook interleaving patterns presented in Section~\ref{sec:codebook_patterns}. Each time step $t_1, t_2, \ldots, t_n$
    is composed of 4 quantized values (corresponding to $k_1, \ldots, k_4$). When doing autoregressive modelling, we can flatten or interleave them in various ways, resulting in a new sequence with 4 parallel streams and steps $s_1, s_2, \ldots, s_m$. The total number of sequence steps $S$ depends on the pattern and original number of steps $T$. $0$ is a special token indicating empty positions in the pattern. \vspace{-0.2cm}
    }
    \label{fig:pattern}
\end{figure*}

\looseness=-1
In this work, we introduce \method, a simple and controllable music generation model, which is able to generate high-quality music given textual description. We propose a general framework for modeling multiple parallel streams of acoustic tokens, which serves as a generalization of previous studies (see~\cref{fig:pattern}). We show how this framework allows to extend generation to stereo audio at no extra computational cost. To improve controllability of the generated samples, we additionally introduce unsupervised melody conditioning, which allows the model to generate music that matches a given harmonic and melodic structure. We conduct an extensive evaluation of \method and show the proposed method is superior to the evaluated baselines by a large margin, with a subjective rating of 84.8 out of 100 for \method against 80.5 for the best baseline. We additionally provide an ablation study which sheds light on the importance of each of the components on the overall model performance. Lastly, human evaluation suggests that \method yields high quality samples which are better melodically aligned with a given harmonic structure, while adhering to a textual description. 

\newparagraph{Our contribution:} (i) We introduce a simple and efficient model to generate high quality music at 32 kHz. We show that \method can generate consistent music with a single-stage language model through an efficient codebook interleaving strategy. (ii) We present a single model to perform both text and melody-conditioned generation and demonstrate that the generated audio is coherent with the provided melody and faithful to the text conditioning information. (iii) We provide extensive objective and human evaluations on the key design choices behind our method.


\vspace{-0.2cm}
\section{Method}\label{sec:method}
\vspace{-0.1cm}

\method consists in an autoregressive transformer-based decoder~\citep{attention}, conditioned on a text or melody representation. The (language) model is over the quantized units from an \encodec~\citep{defossez2022highfi} audio tokenizer, which provides high fidelity reconstruction from a low frame rate discrete representation. Compression models such as \citep{defossez2022highfi, zeghidour2021soundstream} employ Residual Vector Quantization (RVQ) which results in several parallel streams. Under this setting, each stream is comprised of discrete tokens originating from different learned codebooks. Prior work, proposed several modeling strategies to handle this issue~\citep{kharitonov2022text, agostinelli2023musiclm, wang2023neural}. In this work, we introduce a novel modeling framework, which generalizes to various codebook interleaving patterns, and we explore several variants. Through patterns, we can leverage the internal structure of the quantized audio tokens. Finally, \method supports conditional generation based on either text or melody. 

\subsection{Audio tokenization}
\label{sec:audio_tokenization}

\looseness=-1
We use \encodec ~\citep{defossez2022highfi}, a convolutional auto-encoder with a latent space quantized using Residual Vector Quantization (RVQ)~\citep{zeghidour2021soundstream}, and an adversarial reconstruction loss. Given a reference audio random variable $X \in \real^{d \cdot f_s}$ with $d$ the audio duration and $f_s$ the sample rate, \encodec encodes it into a continuous tensor with a frame rate $f_r \ll f_s$. This representation is then quantized into $Q \in \{1, \ldots, M\}^{d \cdot f_r \times K}$, with $K$ being the number of codebooks used in RVQ and $M$ being the codebook size. Notice, after quantization we are left with $K$ parallel discrete tokens sequences, each of length $T = d \cdot f_r$, representing the audio sample.  In RVQ, each quantizer encodes the quantization error left by the previous quantizer, thus quantized values for different codebooks are in general not independent, and the first codebook is the most important one.

\subsection{Codebook interleaving patterns (see~\cref{fig:pattern})}
\label{sec:codebook_patterns}

\newparagraph{Exact flattened autoregressive decomposition.} An autoregressive model requires a discrete random sequence $U \in \{1, \ldots, M\}^{S}$ with $S$ the sequence length.
By convention, we will take $U_0 = 0$, a deterministic special token indicating the beginning of the sequence.
We can then model the distribution
\begin{equation}
\label{eq:autoregressive_model}
    \forall t > 0, p_t\left(U_{t-1}, \ldots, U_0\right) \triangleq \proba{U_t | U_{t-1}, \ldots, U_0}.
\end{equation}
Let us build a second sequence of random variables $\tilde{U}$ using the auto-regressive density $p$, e.g. we define recursively $\tilde{U}_0 = 0$, and for all $t > 0$,
\begin{equation}
\label{eq:autoregressive_generation}
    \forall t > 0, \proba{\tilde{U}_t | \tilde{U}_{t-1} \ldots, \tilde{U}_0} = p_t\left(\tilde{U}_{t-1}, \ldots, \tilde{U}_0\right).
\end{equation}
Then, we immediately have that $U$ and $\tilde{U}$ follow the same distribution. This means that if we can
fit a perfect estimate $\hat{p}$ of $p$ with a deep learning model, then we can fit exactly the distribution of $U$.

As stated before, the main issue with the representation $Q$ we obtained from the \encodec model is that there are $K$ codebooks for each time step. One solution would be to flatten 
out $Q$, thus taking $S = d \cdot f_r \cdot K$, e.g. first predicting the first codebook of the first time step, then the second codebook of the first time step, etc.
Then, using eq. \eqref{eq:autoregressive_model} and eq. \eqref{eq:autoregressive_generation}, we could theoretically fit an exact model of the distribution of $Q$.
The downside however is the increased complexity, with part of the gain coming from the lowest sample rate $f_r$ being lost.

More than one possible flattening exists, and not all the $\hat{p}_t$ functions need to be estimated through a single model. For instance, MusicLM~\citep{agostinelli2023musiclm}
uses two models, one modeling the flattened first $K/2$ codebooks, and a second one the other $K/2$ flattened codebooks, conditioned
on the decision of the first model. In that case, the number of autoregressive steps is still $d f_r \cdot K$.

\newparagraph{Inexact autoregressive decomposition.} Another possibility is to consider an autoregressive decomposition, where some codebooks are predicted in \emph{parallel}. For instance, let us define another sequence with $V_0 = 0$
and for all $t \in \{1, \ldots, T\}$, $k \in \{1, \ldots, K\}$, $V_{t, k} = Q_{t, k}$. When dropping the codebook index $k$, e.g. $V_t$, we mean the concatenation of all the codebooks at time $t$.
\begin{equation}
\label{eq:inexact_autoregressive_model}
    p_{t, k}\left(V_{t-1}, \ldots, V_0\right) \triangleq  \proba{V_{t, k} | V_{t-1}, \cdot, \ldots, V_0}.
\end{equation}
Let's define again recursively $\tilde{V}_0 = 0$ and for all $t > 0$,
\begin{equation}
\label{eq:inexact_autoregressive_generation}
    \forall t > 0, \forall k, \proba{\tilde{V}_{t, k}} = p_{t, k}\left(\tilde{V}_{t-1}, \ldots, \tilde{V}_0\right).
\end{equation}
Unlike in \eqref{eq:autoregressive_generation}, we no longer have in the general case that $\tilde{V}$ follows the same distribution as $V$,
even assuming we have access to the exact distribution $p_{t, k}$. 
In fact, we would only have a proper generative model if for all $t$, $(V_{t, k})_k$ are independent conditionally on $V_{t-1}, \ldots, V_0$.
As $t$ increases, the errors will compound and the two distributions can grow further apart.
Such a decomposition is \emph{inexact}, but allows to keep the original frame rate which can considerably speed up training and inference, 
especially for long sequences.

\newparagraph{Arbitrary codebook interleaving patterns.} In order to experiment with various such decompositions, and measure exactly the impact of using an inexact decomposition, we introduce \emph{codebook interleaving patterns}. Let us consider $\Omega = \{(t, k): \{1, \ldots, d \cdot f_r\}, k \in \{1, \ldots, K\}\}$ be the set of all pairs of time steps and codebook indexes. A codebook pattern is a sequence $P = (P_0, P_1, P_2, \ldots, P_S)$, with $P_0 = \emptyset$, and for all $0 < s \leq S$, $P_s \subset \Omega$, such that $P$ is partition of $\Omega$.  We model $Q$ by predicting in parallel all the positions in $P_s$, conditionally on all the positions in $P_0, P_1, \ldots, P_{s-1}$. Pragmatically, we restrict ourselves to patterns where each codebook index appears at most once in any of the $P_s$.

We can now easily define a  number of decompositions, for instance the ``parallel'' pattern given by
\begin{equation}
    P_s = \{(s, k): k \in \{1, \ldots, K\}\}.
\end{equation}

It is also possible to introduce a ``delay'' between the codebooks, as in~\citet{kharitonov2022text}, e.g.,
\begin{equation}
    P_s = \{(s - k + 1, k): k \in \{1, \ldots, K\}, s - k \geq 0\}.
\end{equation}

Through empirical evaluations, we show the benefits and drawbacks of various codebook patterns, shedding light on the importance of exact modeling of the parallel codebook sequences. 

\subsection{Model conditioning}\label{sec:conditioning}

\newparagraph{Text conditioning.} Given a textual description matching the input audio $X$, we compute a conditioning tensor $C \in \real^{T_C\times D}$ with $D$
being the inner dimension used in the autoregressive model. Generally, there are three main approaches for representing text for conditional audio generation. \citet{kreuk2022audiogen} proposed using a pretrained text encoder, specifically T5~\citep{raffel2020t5}. \citet{chung2022scaling} show that using instruct-based language models provide superior performance. Lastly,~\citet{agostinelli2023musiclm, liu2023audioldm, huang2023make, sheffer2023hear} claimed that joint text-audio representation, such as CLAP~\citep{laionclap2023}, provides better-quality generations. We experiment with all of the above, respectively: T5 encoder, FLAN-T5, and CLAP.

\newparagraph{Melody conditioning.} While text is the prominent approach in conditional generative models nowadays, a more natural approach for music is conditioning on a melodic structure from another audio track or even whistling or humming. Such an approach also allows for an iterative refinement of the model's output. To support that, we experiment with controlling the melodic structure via jointly conditioning on the input's chromagram and text description. In preliminary experiments, we observed that conditioning on the raw chromagram often led to reconstructing the original sample, resulting in overfitting. To reduce it, we introduce an information bottleneck by choosing the dominant time-frequency bin in each time step. While a similar capability was shown in \citet{agostinelli2023musiclm}, the authors used supervised proprietary data, which is tedious and costly to collect. In this work, we take an unsupervised approach, eliminating the requirement for supervised data.

\subsection{Model architecture}
\label{sec:architecture}

\newparagraph{Codebook projection and positional embedding.}
Given a codebook pattern, only some codebooks are present at each pattern step $P_s$. We retrieve from $Q$ the values corresponding to the indices in $P_s$. As noted in~\cref{sec:codebook_patterns}, each codebook is present at most once in $P_s$ or not at all. If it is present, we use a learned embedding table with $N$ entries and dimension $D$ to represent the associated value from $Q$. Otherwise, we use a special token indicating its absence. We sum the contribution from each codebook after this transformation. As $P_0 = \emptyset$, the first input is always the sum of all the special tokens. Finally, we sum a sinusoidal embedding to encode the current step $s$~\citep{attention}.

\newparagraph{Transformer decoder.} The input is fed into a transformer with $L$ layers and a dimension $D$. Each layer consists of a causal self-attention block. We then use a cross-attention block that is fed with the conditioning signal $C$. When using melody conditioning, we instead provide the conditioning tensor $C$ as a prefix to the transformer input. The layer ends with a fully connected block consisting of a linear layer from $D$ to $4 \cdot D$ channels, a ReLU, and a linear layer back to $D$ channels. The attention and fully connected blocks are wrapped with a residual skip connection. Layer normalization~\citep{ba2016layer} is applied to each block before being summed with the residual skip connection (``pre-norm''). 

\newparagraph{Logits prediction.} The output from the transformer decoder at pattern step $P_s$ is transformed into logits prediction for the values of $Q$ taken at the indices given by $P_{s+1}$. Each codebook is present at most once in $P_{s + 1}$. If a codebook is present, the logits prediction is obtained by applying a codebook specific linear layer from $D$ channels to $N$.


\section{Experimental setup}\label{sec:exp}

\subsection{Models and hyperparameters}\label{sec:hyperparams}

\newparagraph{Audio tokenization model.} We use a non-causal five layers \encodec model for $32$ kHz monophonic audio with a stride of $640$, resulting in a frame rate of $50$ Hz, and an initial hidden size of $64$, doubling at each of the model's five layers. The  embeddings are quantized with a RVQ with four quantizers, each with a codebook size of $2048$. We follow \citet{defossez2022highfi} to train the model on one-second audio segments cropped at random in the audio sequence. 

\newparagraph{Transformer model.} We train autoregressive transformer models at different sizes: $300$M, $1.5$B, $3.3$B parameters. We use a memory efficient Flash attention~\citep{dao2022flashattention} from the xFormers package~\citep{xFormers2022} to improve both speed and memory usage with long sequences. We study the impact of the size of the model in \cref{sec:res}. We use the $300$M-parameter model for all of our ablations. We train on $30$-second audio crops sampled at random from the full track. We train the models for $1$M steps with the AdamW optimizer~\citep{loshchilov2017decoupled}, a batch size of $192$ examples, $\beta_1=0.9$, $\beta_2=0.95$, a decoupled weight decay of $0.1$ and gradient clipping of $1.0$. We further rely on D-Adaptation based automatic step-sizes~\citep{defazio2023learning} for the $300$M model as it improves model convergence but showed no gain for the bigger models. We use a cosine learning rate schedule with a warmup of $4000$ steps. Additionally, we use an exponential moving average with a decay of $0.99$. We train the $300$M, $1.5$B and $3.3$B parameter models, using respectively $32$, $64$ and $96$ GPUs, with mixed precision. More specifically, we use float16 as bfloat16 was leading to instabilities in our setup. Finally, for sampling, we employ top-k sampling~\citep{fan2018hierarchical} with keeping the top $250$ tokens and a temperature of $1.0$.

\newparagraph{Text preprocessing.}~\citet{kreuk2022audiogen} proposed a text normalization scheme, in which stop words are omitted and the remaining text is lemmatized. We denote this method by text-normalization. When considering musical datasets, additional annotations tags such as musical key, tempo, type of instruments, etc. are often available. We also experiment with concatenating such annotations to the text description. We denote this approach by condition-merging. Finally, we explored using word dropout as another text augmentation strategy. For the final models, we used condition-merging with a probability of $0.25$. Upon merging, we apply a text description dropout with a probability of $0.5$. We use a word dropout with a probability of $0.3$ on the resulting text. 
A full comparison of the different text preprocessing strategies can be found in~\cref{sec:text_aug}.  

\newparagraph{Codebook patterns and conditioning.} We use the ``delay'' interleaving pattern from \cref{sec:codebook_patterns}, This translates $30$ seconds of audio into $1500$ autoregressive steps.
For text conditioning, we use the T5~\citep{raffel2020t5} text encoder, optionally with the addition of the melody conditioning presented in Section~\ref{sec:conditioning}.  
We also experiment with FLAN-T5~\citep{chung2022scaling}, and CLAP~\citep{laionclap2023} and compare the performance of \method using each of these text encoders in the ~\cref{sec:text_aug}. For melody conditioning, we compute the chromagrams with a window size of $2^{14}$ and a hop size of $2^{12}$. Using a large window prevents the model from recovering fine temporal details. We further quantize the chromagram by taking the argmax at each time step. We follow a similar approach to \citet{kreuk2022audiogen} and implement classifier-free guidance when sampling from the model's logits. Specifically, during training we drop the condition with a probability of $0.2$ and during inference we use a guidance scale of $3.0$.

\subsection{Datasets}\label{sec:datasets}

\newparagraph{Training datasets.}
We use 20K hours of licensed music to train \method. Specifically, we rely on an internal dataset of 10K high-quality music tracks, and on the ShutterStock and Pond5 music data collections\footnote{\href{https://www.shutterstock.com/music}{www.shutterstock.com/music} and \href{https://www.pond5.com}{www.pond5.com}} with respectively 25K and 365K instrument-only music tracks. All datasets consist of full-length music sampled at 32 kHz with metadata composed of a textual description and information such as the genre, BPM, and tags.
We downmix the audio to mono unless stated otherwise.

\newparagraph{Evaluation datasets.} For the main results and comparison with prior work, we evaluate the proposed method on the MusicCaps benchmark \citep{agostinelli2023musiclm}. MusicCaps is composed of $5.5$K samples (ten-second long) prepared by expert musicians and a $1$K subset balanced across genres. We report objective metrics on the unbalanced set, while we sample examples from the genre-balanced set for qualitative evaluations. For melody evaluation and the ablation studies, we use samples from an in-domain held out evaluation set of $528$ music tracks, with no artist overlap with the training set.

\begin{table}[t!]
  \caption{Text-to-Music generation.\label{tab:ttm} We compare objective and subjective metrics for \method against a number of baselines. We report both mean and CI95 scores. The Mousai model is retrained on the same dataset, while for MusicLM we use the public API for human studies. We report the original FAD on MusicCaps for Noise2Music and MusicLM. ``\method w. random melody'' refers to \method trained with chromagram and text. At evaluation time, we sample the chromagrams at random from a held-out set.}  
  \centering
  \resizebox{\columnwidth}{!}{
  \begin{tabular}{lrrr|cc}
    \toprule
       \multicolumn{1}{c}{}& \multicolumn{5}{c}{\footnotesize\textsc{MusicCaps} Test Set}\\ \cmidrule{2-6}
    \textsc{Model}       & \textsc{Fad}$_{vgg} \downarrow$      & \textsc{Kl} $\downarrow$ & \textsc{Clap}$_{scr} \uparrow$ & \textsc{Ovl.} $\uparrow$ & \textsc{Rel.} $\uparrow$\\
    \midrule
    Riffusion               & 14.8      & 2.06  & 0.19 & 79.31\pmr{1.37} & 74.20\pmr{2.17} \\
    Mousai                  & 7.5       & 1.59  & 0.23 & 76.11\pmr{1.56} & 77.35\pmr{1.72} \\
    MusicLM                 & 4.0       & -     & -  & 80.51\pmr{1.07}   & 82.35\pmr{1.36}\\
    Noise2Music             & \textbf{2.1}       & -   & -  & -   & -\\
    \midrule
    \method w.o melody (300M)  & 3.1 & 1.28  & 0.31 & 78.43\pmr{1.30}  & 81.11\pmr{1.31}\\
    \method w.o melody (1.5B) & 3.4    & 1.23  & \textbf{0.32} & 80.74\pmr{1.17}  & \textbf{83.70}\pmr{1.21}\\
    \method w.o melody (3.3B) & 3.8    & \textbf{1.22}  & 0.31 & \textbf{84.81}\pmr{0.95} & 82.47\pmr{1.25}\\
    \method w. random melody (1.5B)    & 5.0   & 1.31 & 0.28  & 81.30\pmr{1.29}  & 81.98\pmr{1.79}\\
    \bottomrule
  \end{tabular}}
\end{table}

\subsection{Evaluation}

\newparagraph{Baselines.} We compare \method to two baselines for text-to-music generation: Riffusion~\citep{forsgrenriffusion} and Mousai~\citep{schneider2023mo}. We use the open source Riffusion model to run inference \footnote{Using riffusion-model-v1 from \href{https://github.com/riffusion/riffusion-app}{github.com/riffusion/riffusion-app} (on May 10, 2023)}. For Mousai, we train a model using our dataset for a fair comparison, using the open source implementation provided by the authors\footnote{Implementation from \href{https://github.com/archinetai/audio-diffusion-pytorch}{github.com/archinetai/audio-diffusion-pytorch} (March 2023)}. 
Additionally, when possible, we compare to MusicLM~\citep{agostinelli2023musiclm} and Noise2Music~\citep{huang2023noise2music}. 

\newparagraph{Evaluation metrics.} We evaluate the proposed method using objective and subjective metrics. For the objective methods, we use three metrics: the Fr\'echet Audio Distance (FAD), the Kullback-Leiber Divergence (KL) and the CLAP score (CLAP). 
We report the FAD~\citep{kilgour2018fr} using the official implementation in Tensorflow with the VGGish model~\footnote{\href{https://github.com/google-research/google-research/tree/master/frechet_audio_distance}{github.com/google-research/google-research/tree/master/frechet\_audio\_distance}}. 
A low FAD score indicates the generated audio is plausible. 
Following~\citet{kreuk2022audiogen}, we use a state-of-the-art audio classifier trained for 
classification on AudioSet~\citep{koutini2021efficient} to compute the KL-divergence over the probabilities of the labels between the original and the generated music. The generated music is expected to share similar concepts with the reference music when the KL is low.  Last, the CLAP score~\citep{laionclap2023, huang2023make} is computed between the track description and the generated audio to quantify audio-text alignment, using the official pretrained CLAP model~\footnote{\href{https://github.com/LAION-AI/CLAP}{https://github.com/LAION-AI/CLAP}}. 

For the human studies, we follow the same setup as in~\citet{kreuk2022audiogen}. We ask human raters to evaluate two aspects of the audio samples (i) overall quality (\textsc{Ovl}), and (ii) relevance to the text input (\textsc{Rel}). For the overall quality test, raters were asked to rate the perceptual quality of the provided samples in a range of $1$ to $100$. For the text relevance test, raters were asked to rate the match between audio and text on a scale of $1$ to $100$. Raters were recruited using the Amazon Mechanical Turk platform. We evaluate randomly sampled files, where each sample was evaluated by at least $5$ raters. We use the CrowdMOS package\footnote{\href{http://www.crowdmos.org/download/}{http://www.crowdmos.org/download/}} to filter noisy annotations and outliers. We remove annotators who did not listen to the full recordings, annotators who rate the reference recordings less than $85$, and the rest of the recommended recipes from  CrowdMOS~\citep{ribeiro2011crowdmos}. For fairness, all samples are normalized at ${-}14$dB LUFS~\citep{itur2012algorithms}.

\section{Results}
\label{sec:res}
We start by presenting results of the proposed method on the task of text-to-music generation and compare \method to prior work in the field. Next, we evaluate the ability of the proposed method to generate music conditioned on melodic features. We further show how to simply extend our codebook patterns for stereo audio generation. We conclude with an ablation study. 
Music samples, code, and models are available at \href{https://github.com/facebookresearch/audiocraft}{github.com/facebookresearch/audiocraft}.

\subsection{Comparison with the baselines}
\label{sec:comp_baselines}
Table~\ref{tab:ttm} presents the comparison of the proposed method against Mousai, Riffusion,  MusicLM, and Noise2Music. As there is no official implementation of Noise2Music, nor pre-trained model, we report only FAD on MusicCaps, which is reported in the Noise2Music manuscript. Similarly, MusicLM implementation is not public. We use the MusicLM public demo\footnote{\href{https://blog.google/technology/ai/musiclm-google-ai-test-kitchen/}{https://blog.google/technology/ai/musiclm-google-ai-test-kitchen/}} for our subjective tests while reporting the FAD as reported by the authors. While the original MusicLM model is trained on data with vocals, the model behind the API is instrument-only. For the human study, we restrict ourselves to $40$ instrument-only samples from MusicCaps. To prevent leakage in \method trained with chromagram, we sample chromagrams at random from a held-out set during test time. 

Results suggest that \method performs better than the evaluated baselines as evaluated by human listeners, both in terms of audio quality and adherence to the provided text description. Noise2Music performs the best in terms of FAD on MusicCaps, followed by \method trained with text conditioning. Interestingly, adding a melody conditioning degrades the objective metrics, however, it does not significantly affect human ratings, while still being superior to the evaluated baselines. 

We notice that for the worst rated model, the FAD is correlated with the overall subjective rating, but it is not
the case for the best rated models. We noticed that a large number of samples in MusicCaps~\citep{agostinelli2023musiclm} contains a description stating that the recording is "noisy". It is possible that due to those noisy samples, improvements
in the quality of the generated audio might deteriorate the FAD on MusicCaps once a certain quality threshold is achieved.

\subsection{Melody evaluation}
We evaluate \method, conditioned jointly on textual and melodic representations, using objective and subjective metric on the held out evaluation set.  For the objective evaluation, we introduce a new metric: chroma cosine-similarity, which measures the average cosine-similarity between frames corresponding to the same time steps, taken from the quantized chroma of the reference and the generated samples. We evaluate using $1000$ randomly sampled files from a held-out set. To better evaluate the relation between the conditioned melody to the generated music, we introduce another human study. To that end, we present human raters with a reference musical piece, followed by a set of generated pieces. For each generated sample, the listeners are asked to rate how well the melody of the generated piece matches that of the reference on a scale of $1$ to $100$. We use $40$ samples of $10$ seconds at random from the held-out set. Results are reported in \cref{tab:melody}. Results suggest that \method trained with chromagram conditioning successfully generates music that follows a given melody. Thus, allowing for better control over the generated output. Interestingly, \method is robust to dropping the chroma at inference time with both \textsc{Ovl.} and \textsc{Rel.} staying roughly the same.

\begin{table}[t!]
  \caption{We report cosine similarity between reference and generated melody (\textsc{Sim.}) and subjective metrics including alignment with the melody (\textsc{Mel.}). All results are reported for \method 1.5B.} \label{tab:melody}  
  \centering
  \resizebox{0.85\columnwidth}{!}{%
  \begin{tabular}{l|l|c|ccc}
    \toprule
        \multicolumn{2}{c}{}& \multicolumn{4}{c}{\footnotesize In Domain Test Set}\\ \cmidrule{3-6}
    \textsc{Train condition} & \textsc{Test condition} & \textsc{Sim.} $\uparrow$ & \textsc{Mel.} $\uparrow$  & \textsc{Ovl.} $\uparrow$ & \textsc{Rel.} $\uparrow$ \\
    \midrule
    Text & Text & 0.10 & 64.44\pmr{0.83} & 82.18\pmr{1.21} & 81.54\pmr{1.22} \\
    Text+Chroma  & Text & 0.10 & 61.89\pmr{0.96} & 81.65\pmr{1.13} & \textbf{82.50}\pmr{0.98} \\
    Text+Chroma & Text+Chroma & \textbf{0.66} & \textbf{72.87\pmr{0.93}} & \textbf{83.94}\pmr{1.99} & 80.28\pmr{1.06} \\
    \bottomrule
  \end{tabular}}
\end{table}

\vspace{-0.2cm}
\subsection{Fine-tuning for stereophonic generation}
\vspace{-0.1cm}

\begin{figure}
    \centering
    \begin{subfigure}[t]{0.58\textwidth}
        \includegraphics[width=\textwidth]{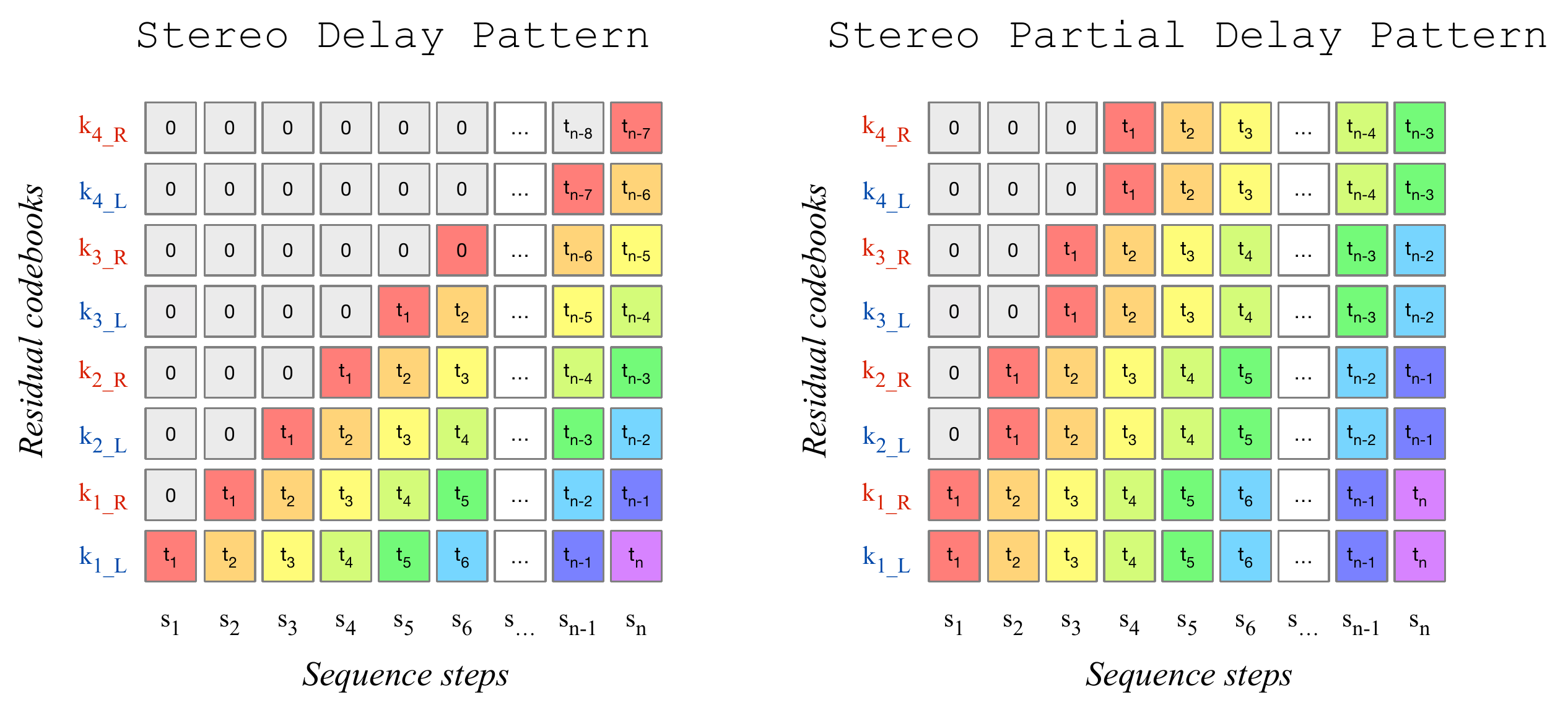}
        \caption{Visualizing codebook patterns for stereo models with two possible interleaving of the left (blue) and right (red) codebooks. For a given codebook index, the ``stereo delay'' pattern uses different delays for the left and right channels,
        while the ``stereo partial delay'' predicts the two channels in parallel.}
        \label{fig:pattern_stereo}
    \end{subfigure}%
    \hfill %
    \begin{subfigure}[t]{0.35\textwidth}
        \centering
            \includegraphics[width=\textwidth]{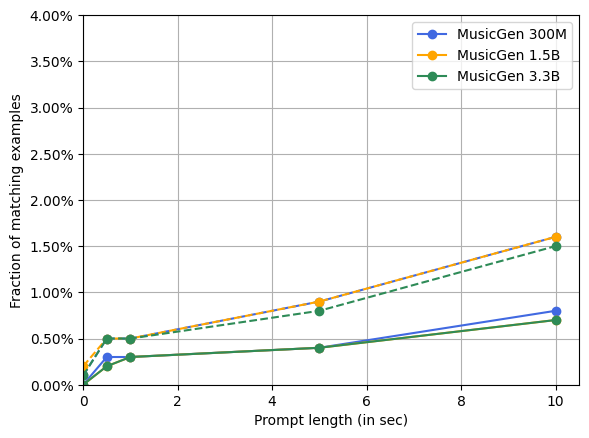}
               \caption{Memorization results for the first codebook tokens on 5-second audio generations, considering exact (solid line) and 80\% partial (dashed line) matches when prompted with extracts of varying duration from the train set.}
            \label{fig:memorization} 
    \end{subfigure}
    \caption{Stereo codebooks (left) and memorization results (right)}
\end{figure}

We experiment with for extending generation to stereophonic data.
We use the same \encodec tokenizer, which we applied independently to the left and right channels, providing $2 \cdot K = 8$ codebooks per frame. Starting from a pre-trained monophonic \method model, we fine tune it for 200K batches the same dataset with stereo audio. We reuse the ``delay'' pattern, with two possible variations: (i) ``stereo delay'' introduces a delay between the left and right channels for the same codebook level, while (ii) ``stereo partial delay'' apply the same delay to the left and right channel codebooks for a given level, as depicted in Figure~\ref{fig:pattern_stereo}.
Note that using this simple strategy, we can generate stereo audio \emph{at no extra computational cost}.
We provide in Table~\ref{tab:stereo} the subjective evaluations for those models. We notice that when downmixing the stereo output to mono, we are almost equivalent in perceived quality to a mono model. Stereo audio was overall rated higher than the mono counterpart, and the ``stereo partial delay'' benefits from a small boost both in overall quality and text relevance compared with the ``stereo delay'' pattern.

\begin{table}
      \caption{Stereophonic Text-to-Music generation.
        EnCodec processes separately the left and right channels, giving us 8 codebooks instead of 4.
        We experiment with two codebook patterns, depicted in Figure~\ref{fig:pattern_stereo}.
        We also measure one of the stereo model after being downmixed to mono. 
        We use a 1.5B \method model conditioned only on text. 
        \label{tab:stereo}}  
      \centering
      \begin{tabular}{ll|cc}
        \toprule
           \multicolumn{2}{c}{}& \multicolumn{2}{c}{\footnotesize\textsc{MusicCaps} Test Set}\\ \cmidrule{3-4}
        \textsc{Cb. Pattern}  & \textsc{Stereo?}   & \textsc{Ovl.} $\uparrow$ & \textsc{Rel.} $\uparrow$\\
        \midrule
        \emph{Mono Delay}  & \crmark & 84.95\pmr{1.60}  & \textbf{80.61}\pmr{1.22} \\
        \emph{Stereo Partial Delay} & \crmark* & 84.49\pmr{1.80}  & 79.39\pmr{1.16}\\
        \emph{Stereo Partial Delay} & \chmark & \textbf{86.73}\pmr{1.06}  & \textbf{80.41}\pmr{1.15}\\
        \emph{Stereo Delay} & \chmark & 85.51\pmr{1.21}  & 78.32 \pmr{1.21}\\
        \bottomrule
      \end{tabular}
      \\
      {\footnotesize *: downmixed to mono after generation.}
\end{table}

\vspace{-0.2cm}
\subsection{Ablation}
\vspace{-0.1cm}

This section provides an ablation study for the different codebook patterns, together with results for model scales and a memorization study. Additionally, we present results for different text augmentation strategies, text encoders, and audio tokenization models in~\cref{app:sec:results}. All ablations are performed using $1$K samples of $30$ seconds, randomly sampled from the held-out evaluation set.

\begin{table}[t!]
  \caption{Codebook patterns. We compare different codebook interleaving patterns on $30$-seconds, audio sequences. The ``flattening'' pattern achieves the best scores. The ``delay'' and ``partial flattening'' patterns achieve similar scores, while ``parallel'' obtains worse scores.\label{tab:cb_strat}}  
  \centering
      \resizebox{0.9\textwidth}{!}{
  \begin{tabular}{l|c|rrr|cc}
    \toprule
             \multicolumn{2}{c}{}& \multicolumn{5}{c}{\footnotesize In Domain Test Set}\\ \cmidrule{3-7}
    \textsc{Configuration} & Nb. steps & \textsc{Fad}$_{\text{vgg}} \downarrow$      & \textsc{Kl} $\downarrow$ & \textsc{Clap}$_{\text{scr}} \uparrow$  & \textsc{Ovl.} $\uparrow$ & \textsc{Rel.} $\uparrow$\\
    \midrule
    Delay & 1500 & 0.96 & 0.52 & 0.35 & \textbf{79.69}\pmr{1.46} & 79.67\pmr{1.41} \\
    Partial delay & 1500 & 1.51 & 0.54 & 0.32 & 79.13\pmr{1.56} & 79.67\pmr{1.46}\\
    Parallel & 1500 & 2.58 & 0.62 & 0.27 & 72.21\pmr{2.49} & 80.30\pmr{1.43} \\
    \midrule
    Partial flattening & 3000 & 1.32 & 0.54 & 0.34 & 78.56\pmr{1.86} & 79.18\pmr{1.49} \\
    Coarse first & 3000 & 1.98 & 0.56 & 0.30 & 74.42\pmr{2.28} & 76.55\pmr{1.67} \\
    \midrule
    Flattening & 6000 & \textbf{0.86} & \textbf{0.51} & \textbf{0.37} & \textbf{79.71}\pmr{1.58} & \textbf{82.03}\pmr{1.1} \\
    \bottomrule
  \end{tabular}
  }
  \vspace{-0.1cm}
\end{table}

\newparagraph{The effect of the codebook interleaving patterns.} We evaluate various codebook patterns using the framework from~\cref{sec:codebook_patterns}, with $K=4$, given by the audio tokenization model. \cref{tab:ttm} reports results with the ``delay'' pattern. The ``partial delay'' consists in delaying by the same amount the codebooks 2, 3, and 4.  The ``parallel'' pattern predicts all the codebooks from the same time step in parallel. The ``coarse first'' pattern first predicts codebook 1 for all steps, then predicts in parallel codebooks 2, 3, and 4. Thus, this pattern has twice the steps compared to other patterns. ``Partial flattening'' is similar, but instead of sampling first codebook 1 for all steps, it interleaves them with the parallel sampling of codebooks 2, 3, and 4. Finally, the ``flattening'' pattern consists in flattening all the codebooks, similar to MusicLM~\citep{agostinelli2023musiclm}. We report objective and subjective evaluations in \cref{tab:cb_strat}. While flattening improves generation, it comes at a high computational cost and  similar performance can be reached for a fraction of this cost using a simple delay approach.

\newparagraph{The effect of model size.} In \cref{tab:model_scale} we report results for different model sizes, namely 300M, 1.5B, and 3.3B parameter models. As expected, scaling the model size results in better scores, however this comes at the expense of longer training and inference time. Regarding subjective evaluations, the overall quality is optimal at 1.5B, but a larger model can better understand the text prompt.

\newparagraph{Memorization experiment.} Following \citep{agostinelli2023musiclm}, we analyze the memorization abilities of \method. We only consider the first stream of codebooks from \method as it contains the coarser-grain information.
We randomly select $N=20,000$ examples from our training set and for each one we feed the model with a prompt of EnCodec codebooks corresponding to the original audio and the conditioning information. We generate a continuation of 250 audio tokens (5 second-long audio) using greedy decoding. We report exact matches as the fraction of examples for which the generated audio tokens and source audio tokens are identical over the whole sequence. In addition, we report partial matches as the fraction of the training examples for which the generated and source sequences have at least $80\%$ of the audio tokens matching. We present the memorization results for the different model sizes when varying the length of the audio prompt in Figure~\ref{fig:memorization}. 

\begin{table}[t!]
  \caption{Model scale.\label{tab:model_scale} We compare 3 scales for our method, and evaluate it on an internal test set to limit the impact of the out of domain prediction issues we observed with MusicCaps. 
  In terms of objective metrics we observe a continuous improvement of the metrics, although subjective quality stop improving at 1.5B parameters. A 3.3B model however seems to better fit the text prompt.}  
  \centering
    \resizebox{0.95\textwidth}{!}{
  \begin{tabular}{rrr|c|cccc|cc}
    \toprule
         \multicolumn{4}{c}{}& \multicolumn{6}{c}{\footnotesize In Domain Test Set}\\
     \cmidrule{5-10}
    Dim. & Heads & Depth & \# Param. & \textsc{Ppl}$ \downarrow$ & \textsc{Fad}$_{\text{vgg}} \downarrow$      & \textsc{Kl} $\downarrow$ & \textsc{Clap}$_{\text{scr}} \uparrow$ & \textsc{Ovl.} $\uparrow$ & \textsc{Rel.} $\uparrow$ \\
    \midrule
    1024  & 16 & 24 & 300M  & 56.1 & 0.96      & 0.52  & 0.35 & 78.3\pmr{1.4} & 82.5 \pmr{1.6} \\
    1536 & 24 & 48 & 1.5B   & 48.4 & 0.86      & \textbf{0.50}  & 0.35  & \textbf{81.9}\pmr{1.4} & 82.9\pmr{1.5}\\    
    2048 & 32 & 48 & 3.3B   & \textbf{46.1} & \textbf{0.82}      & \textbf{0.50}  & \textbf{0.36}   & 79.2\pmr{1.3}  &  \textbf{83.5}\pmr{1.3} \\
    \bottomrule
  \end{tabular}}
  \vspace{-0.2cm}
\end{table}

\vspace{-0.2cm}
\section{Related work}\label{sec:rel}
\vspace{-0.1cm}

\newparagraph{Audio representation.} In recent years, the prominent approach is to represent the music signals in a compressed representation, discrete or continuous, and apply a generative model on top of it.~\citet{lakhotia2021generative} proposed quantizing speech representation using k-means to construct speech language models. Recently,~\citet{defossez2022highfi, zeghidour2021soundstream} proposed to apply a VQ-VAE directly on the raw waveform using residual vector quantization. Later, several studies used such representation for text-to-audio generation. Next, we discuss the recent research in audio generation.

\newparagraph{Music generation.} Music generation has been long studied under various setups.~\citet{dong2018musegan} proposed a GAN-based approach for symbolic music generation.~\citet{bassan2022unsupervised} proposed an unsupervised segmentation for symbolic music which can be later used for generation.~\citet{ycart2017study} proposed modeling polyphonic music modeling using recurrent neural networks. ~\citet{ji2020comprehensive} conducted a comprehensive survey therein for deep learning methods for music generation.

\citet{dhariwal2020jukebox} proposed representing music samples in multiple streams of discrete representations using a hierarchical VQ-VAE. Next, two sparse transformers applied over the sequences to generate music.~\citet{gan2020foley} proposed generating music for a given video, while predicting its midi notes. Recently,~\citet{agostinelli2023musiclm} proposed representing music using multiple streams of ``semantic tokens'' and ``acoustic tokens''. Then, they applied a cascade of transformer decoders conditioned on a textual-music joint representation~\citep{huang2022mulan}.~\citet{donahue2023singsong} followed a similar modeling approach, but for the task of singing-to-accompaniment generation. 

\looseness=-1
An alternative approach is using diffusion models.~\citet{schneider2023mo, huang2023noise2music, maina2023msanii, forsgrenriffusion} proposed using a latent diffusion model for the task of text-to-music.~\citet{schneider2023mo} proposed using diffusion models for both audio encoder-decoder and latent generation.~\citet{huang2023noise2music} proposed a cascade of diffusion model to generate audio and gradually increase its sampling rate.~\citet{forsgrenriffusion} proposed fine-tuning Stable Diffusion~\citet{rombach2022high} using spectrograms to generate five-second segments, then, using image-to-image mapping and latent interpolation to generate long sequences.

\newparagraph{Audio generation.} Several studies were proposed for text-to-audio (environmental sounds) generation.~\citet{yang2022diffsound} proposed representing audio spectrograms using a VQ-VAE, then applying a discrete diffusion model conditioned on textual CLIP embeddings for the generation part~\citep{radford2021learning}.~\citet{kreuk2022audiogen} proposed applying a transformer language model over discrete audio representation, obtained by quantizing directly time-domain signals using \encodec~\citep{defossez2022highfi}.~\citet{sheffer2023hear} followed a similar approach to \citet{kreuk2022audiogen} for image-to-audio generation.~\citet{huang2023make, liu2023audioldm} proposed using latent diffusion models for the task of text-to-audio, while extending it to various other tasks such as inpainting, image-to-audio, etc.
\vspace{-0.15cm}
\section{Discussion}\label{sec:dis}
\vspace{-0.15cm}

\looseness=-1
We introduced \method, a state-of-the-art single stage controllable music generation model that can be conditioned on text and melody. We demonstrated that simple codebook interleaving strategies can be used to achieve high quality generation, even in stereo, while reducing the number of autoregressive time steps compared to the flattening approach. We provided a comprehensive study of the impact of model sizes, conditioning methods, and text pre-processing techniques. We also introduced a simple chromagram-based conditioning for controlling the melody of the generated audio.

\newparagraph{Limitations}
Our simple generation method does not allow us to have fine-grained control over adherence of the generation to the conditioning, we rely mostly on CF guidance. Also, while it is relatively straightforward to do data augmentation for text conditioning, conditioning on audio warrants further research on data augmentation, types and amount of guidance.

\newparagraph{Broader impact.} 
Large scale generative models present ethical challenges. We first ensured that all the data we trained on was covered by legal agreements with the right holders, in particular through an agreement with ShutterStock. A second aspect is the potential lack of diversity in the dataset we used, which contains a larger proportion of western-style music. However, we believe the simplification we operate in this work, e.g., using a single stage language model and a reduced number of auto-regressive steps, can help broaden the applications to new datasets. 
Generative models can represent an unfair competition for artists, which is an open problem. Open research can ensure that all actors have equal access to these models. Through the development of more advanced controls, such as the melody conditioning we introduced, we hope that such models can become useful both to music amateurs and professionals.

\section*{Acknowledgements.} The authors would like to thank Mary Williamson, Rashel Moritz and Joelle Pineau for supporting this project, thank Justin Luk, Prash Jain, Sidd Srinivasan, Rod Duenes, and Philip Woods for the dataset, and thank the xformers team: Daniel Haziza, Francisco Massa, and Michael Ramamonjisoa for the technical discussions.

\newparagraph{Authors note.} This paper was submitted in the wake of the tragic terror attack perpetrated by Hamas on October 7, 2023, which has left the Israeli nation profoundly devastated. The assault greatly impacted us at a personal level, thereby significantly impacting the course of this research work. This paper was finalized while we grieve and mourn our friends and family, under great stress, with scientific considerations being the last thing on our minds. It may contain subtle errors.

In memory of the countless lives shattered by Hamas actions.

\bibliographystyle{unsrtnat}
\bibliography{refs}

\begin{thebibliography}{47}
\providecommand{\natexlab}[1]{#1}
\providecommand{\url}[1]{\texttt{#1}}
\expandafter\ifx\csname urlstyle\endcsname\relax
  \providecommand{\doi}[1]{doi: #1}\else
  \providecommand{\doi}{doi: \begingroup \urlstyle{rm}\Url}\fi

\bibitem[M{\"u}ller(2015)]{muller2015fundamentals}
Meinard M{\"u}ller.
\newblock \emph{Fundamentals of music processing: Audio, analysis, algorithms,
  applications}, volume~5.
\newblock Springer, 2015.

\bibitem[Fedorenko et~al.(2012)Fedorenko, McDermott, Norman-Haignere, and
  Kanwisher]{fedorenko2012sensitivity}
Evelina Fedorenko, Josh~H McDermott, Sam Norman-Haignere, and Nancy Kanwisher.
\newblock Sensitivity to musical structure in the human brain.
\newblock \emph{Journal of neurophysiology}, 108\penalty0 (12):\penalty0
  3289--3300, 2012.

\bibitem[Norman-Haignere et~al.(2019)Norman-Haignere, Kanwisher, McDermott, and
  Conway]{norman2019divergence}
Sam~V Norman-Haignere, Nancy Kanwisher, Josh~H McDermott, and Bevil~R Conway.
\newblock Divergence in the functional organization of human and macaque
  auditory cortex revealed by fmri responses to harmonic tones.
\newblock \emph{Nature neuroscience}, 22\penalty0 (7):\penalty0 1057--1060,
  2019.

\bibitem[Balestriero et~al.(2023)Balestriero, Ibrahim, Sobal, Morcos, Shekhar,
  Goldstein, Bordes, Bardes, Mialon, Tian, et~al.]{balestriero2023cookbook}
Randall Balestriero, Mark Ibrahim, Vlad Sobal, Ari Morcos, Shashank Shekhar,
  Tom Goldstein, Florian Bordes, Adrien Bardes, Gregoire Mialon, Yuandong Tian,
  et~al.
\newblock A cookbook of self-supervised learning.
\newblock \emph{arXiv preprint arXiv:2304.12210}, 2023.

\bibitem[Touvron et~al.(2023)Touvron, Lavril, Izacard, Martinet, Lachaux,
  Lacroix, Rozi{\`e}re, Goyal, Hambro, Azhar, et~al.]{touvron2023llama}
Hugo Touvron, Thibaut Lavril, Gautier Izacard, Xavier Martinet, Marie-Anne
  Lachaux, Timoth{\'e}e Lacroix, Baptiste Rozi{\`e}re, Naman Goyal, Eric
  Hambro, Faisal Azhar, et~al.
\newblock Llama: Open and efficient foundation language models.
\newblock \emph{arXiv preprint arXiv:2302.13971}, 2023.

\bibitem[Tan et~al.(2021)Tan, Qin, Soong, and Liu]{tan2021survey}
Xu~Tan, Tao Qin, Frank Soong, and Tie-Yan Liu.
\newblock A survey on neural speech synthesis.
\newblock \emph{arXiv preprint arXiv:2106.15561}, 2021.

\bibitem[Défossez et~al.(2022)Défossez, Copet, Synnaeve, and
  Adi]{defossez2022highfi}
Alexandre Défossez, Jade Copet, Gabriel Synnaeve, and Yossi Adi.
\newblock High fidelity neural audio compression.
\newblock \emph{arXiv preprint arXiv:2210.13438}, 2022.

\bibitem[Kharitonov et~al.(2022)Kharitonov, Lee, Polyak, Adi, Copet, Lakhotia,
  Nguyen, Riviere, Mohamed, Dupoux, et~al.]{kharitonov2022text}
Eugene Kharitonov, Ann Lee, Adam Polyak, Yossi Adi, Jade Copet, Kushal
  Lakhotia, Tu~Anh Nguyen, Morgane Riviere, Abdelrahman Mohamed, Emmanuel
  Dupoux, et~al.
\newblock Text-free prosody-aware generative spoken language modeling.
\newblock In \emph{Proceedings of the 60th Annual Meeting of the Association
  for Computational Linguistics (Volume 1: Long Papers)}, pages 8666--8681,
  2022.

\bibitem[Kreuk et~al.(2022)Kreuk, Synnaeve, Polyak, Singer, D{\'e}fossez,
  Copet, Parikh, Taigman, and Adi]{kreuk2022audiogen}
Felix Kreuk, Gabriel Synnaeve, Adam Polyak, Uriel Singer, Alexandre
  D{\'e}fossez, Jade Copet, Devi Parikh, Yaniv Taigman, and Yossi Adi.
\newblock Audiogen: Textually guided audio generation.
\newblock \emph{arXiv preprint arXiv:2209.15352}, 2022.

\bibitem[Agostinelli et~al.(2023)Agostinelli, Denk, Borsos, Engel, Verzetti,
  Caillon, Huang, Jansen, Roberts, Tagliasacchi,
  et~al.]{agostinelli2023musiclm}
Andrea Agostinelli, Timo~I Denk, Zal{\'a}n Borsos, Jesse Engel, Mauro Verzetti,
  Antoine Caillon, Qingqing Huang, Aren Jansen, Adam Roberts, Marco
  Tagliasacchi, et~al.
\newblock Musiclm: Generating music from text.
\newblock \emph{arXiv preprint arXiv:2301.11325}, 2023.

\bibitem[Donahue et~al.(2023)Donahue, Caillon, Roberts, Manilow, Esling,
  Agostinelli, Verzetti, Simon, Pietquin, Zeghidour,
  et~al.]{donahue2023singsong}
Chris Donahue, Antoine Caillon, Adam Roberts, Ethan Manilow, Philippe Esling,
  Andrea Agostinelli, Mauro Verzetti, Ian Simon, Olivier Pietquin, Neil
  Zeghidour, et~al.
\newblock Singsong: Generating musical accompaniments from singing.
\newblock \emph{arXiv preprint arXiv:2301.12662}, 2023.

\bibitem[Wang et~al.(2023)Wang, Chen, Wu, Zhang, Zhou, Liu, Chen, Liu, Wang,
  Li, et~al.]{wang2023neural}
Chengyi Wang, Sanyuan Chen, Yu~Wu, Ziqiang Zhang, Long Zhou, Shujie Liu, Zhuo
  Chen, Yanqing Liu, Huaming Wang, Jinyu Li, et~al.
\newblock Neural codec language models are zero-shot text to speech
  synthesizers.
\newblock \emph{arXiv preprint arXiv:2301.02111}, 2023.

\bibitem[Vaswani et~al.(2017)Vaswani, Shazeer, Parmar, Uszkoreit, Jones, Gomez,
  Kaiser, and Polosukhin]{attention}
Ashish Vaswani, Noam Shazeer, Niki Parmar, Jakob Uszkoreit, Llion Jones,
  Aidan~N Gomez, \L~ukasz Kaiser, and Illia Polosukhin.
\newblock Attention is all you need.
\newblock In I.~Guyon, U.~Von Luxburg, S.~Bengio, H.~Wallach, R.~Fergus,
  S.~Vishwanathan, and R.~Garnett, editors, \emph{Advances in Neural
  Information Processing Systems}, volume~30. Curran Associates, Inc., 2017.
\newblock URL
  \url{https://proceedings.neurips.cc/paper_files/paper/2017/file/3f5ee243547dee91fbd053c1c4a845aa-Paper.pdf}.

\bibitem[Zeghidour et~al.(2021)Zeghidour, Luebs, Omran, Skoglund, and
  Tagliasacchi]{zeghidour2021soundstream}
Neil Zeghidour, Alejandro Luebs, Ahmed Omran, Jan Skoglund, and Marco
  Tagliasacchi.
\newblock Soundstream: An end-to-end neural audio codec.
\newblock \emph{IEEE/ACM Transactions on Audio, Speech, and Language
  Processing}, 2021.

\bibitem[Raffel et~al.(2020)Raffel, Shazeer, Roberts, Lee, Narang, Matena,
  Zhou, Li, and Liu]{raffel2020t5}
Colin Raffel, Noam Shazeer, Adam Roberts, Katherine Lee, Sharan Narang, Michael
  Matena, Yanqi Zhou, Wei Li, and Peter~J Liu.
\newblock Exploring the limits of transfer learning with a unified text-to-text
  transformer.
\newblock \emph{The Journal of Machine Learning Research}, 21\penalty0
  (1):\penalty0 5485--5551, 2020.

\bibitem[Chung et~al.(2022)Chung, Hou, Longpre, Zoph, Tay, Fedus, Li, Wang,
  Dehghani, Brahma, et~al.]{chung2022scaling}
Hyung~Won Chung, Le~Hou, Shayne Longpre, Barret Zoph, Yi~Tay, William Fedus,
  Eric Li, Xuezhi Wang, Mostafa Dehghani, Siddhartha Brahma, et~al.
\newblock Scaling instruction-finetuned language models.
\newblock \emph{arXiv preprint arXiv:2210.11416}, 2022.

\bibitem[Liu et~al.(2023)Liu, Chen, Yuan, Mei, Liu, Mandic, Wang, and
  Plumbley]{liu2023audioldm}
Haohe Liu, Zehua Chen, Yi~Yuan, Xinhao Mei, Xubo Liu, Danilo Mandic, Wenwu
  Wang, and Mark~D Plumbley.
\newblock Audioldm: Text-to-audio generation with latent diffusion models.
\newblock \emph{arXiv preprint arXiv:2301.12503}, 2023.

\bibitem[Huang et~al.(2023{\natexlab{a}})Huang, Huang, Yang, Ren, Liu, Li, Ye,
  Liu, Yin, and Zhao]{huang2023make}
Rongjie Huang, Jiawei Huang, Dongchao Yang, Yi~Ren, Luping Liu, Mingze Li,
  Zhenhui Ye, Jinglin Liu, Xiang Yin, and Zhou Zhao.
\newblock Make-an-audio: Text-to-audio generation with prompt-enhanced
  diffusion models.
\newblock \emph{arXiv preprint arXiv:2301.12661}, 2023{\natexlab{a}}.

\bibitem[Sheffer and Adi(2023)]{sheffer2023hear}
Roy Sheffer and Yossi Adi.
\newblock I hear your true colors: Image guided audio generation.
\newblock In \emph{ICASSP 2023-2023 IEEE International Conference on Acoustics,
  Speech and Signal Processing (ICASSP)}, pages 1--5. IEEE, 2023.

\bibitem[Wu* et~al.(2023)Wu*, Chen*, Zhang*, Hui*, Berg-Kirkpatrick, and
  Dubnov]{laionclap2023}
Yusong Wu*, Ke~Chen*, Tianyu Zhang*, Yuchen Hui*, Taylor Berg-Kirkpatrick, and
  Shlomo Dubnov.
\newblock Large-scale contrastive language-audio pretraining with feature
  fusion and keyword-to-caption augmentation.
\newblock In \emph{IEEE International Conference on Acoustics, Speech and
  Signal Processing, ICASSP}, 2023.

\bibitem[Ba et~al.(2016)Ba, Kiros, and Hinton]{ba2016layer}
Jimmy~Lei Ba, Jamie~Ryan Kiros, and Geoffrey~E Hinton.
\newblock Layer normalization.
\newblock \emph{arXiv preprint arXiv:1607.06450}, 2016.

\bibitem[Dao et~al.(2022)Dao, Fu, Ermon, Rudra, and
  R{\'e}]{dao2022flashattention}
Tri Dao, Daniel~Y. Fu, Stefano Ermon, Atri Rudra, and Christopher R{\'e}.
\newblock Flash{A}ttention: Fast and memory-efficient exact attention with
  {IO}-awareness.
\newblock In \emph{Advances in Neural Information Processing Systems}, 2022.

\bibitem[Lefaudeux et~al.(2022)Lefaudeux, Massa, Liskovich, Xiong, Caggiano,
  Naren, Xu, Hu, Tintore, Zhang, Labatut, and Haziza]{xFormers2022}
Benjamin Lefaudeux, Francisco Massa, Diana Liskovich, Wenhan Xiong, Vittorio
  Caggiano, Sean Naren, Min Xu, Jieru Hu, Marta Tintore, Susan Zhang, Patrick
  Labatut, and Daniel Haziza.
\newblock xformers: A modular and hackable transformer modelling library.
\newblock \url{https://github.com/facebookresearch/xformers}, 2022.

\bibitem[Loshchilov and Hutter(2017)]{loshchilov2017decoupled}
Ilya Loshchilov and Frank Hutter.
\newblock Decoupled weight decay regularization.
\newblock \emph{arXiv preprint arXiv:1711.05101}, 2017.

\bibitem[Defazio and Mishchenko(2023)]{defazio2023learning}
Aaron Defazio and Konstantin Mishchenko.
\newblock Learning-rate-free learning by d-adaptation.
\newblock \emph{arXiv preprint arXiv:2301.07733}, 2023.

\bibitem[Fan et~al.(2018)Fan, Lewis, and Dauphin]{fan2018hierarchical}
Angela Fan, Mike Lewis, and Yann Dauphin.
\newblock Hierarchical neural story generation.
\newblock \emph{arXiv preprint arXiv:1805.04833}, 2018.

\bibitem[Forsgren and Martiros()]{forsgrenriffusion}
S~Forsgren and H~Martiros.
\newblock Riffusion-stable diffusion for real-time music generation. 2022.
\newblock \emph{URL https://riffusion. com/about}.

\bibitem[Schneider et~al.(2023)Schneider, Jin, and
  Sch{\"o}lkopf]{schneider2023mo}
Flavio Schneider, Zhijing Jin, and Bernhard Sch{\"o}lkopf.
\newblock Mo$\backslash$\^{} usai: Text-to-music generation with long-context
  latent diffusion.
\newblock \emph{arXiv preprint arXiv:2301.11757}, 2023.

\bibitem[Huang et~al.(2023{\natexlab{b}})Huang, Park, Wang, Denk, Ly, Chen,
  Zhang, Zhang, Yu, Frank, et~al.]{huang2023noise2music}
Qingqing Huang, Daniel~S Park, Tao Wang, Timo~I Denk, Andy Ly, Nanxin Chen,
  Zhengdong Zhang, Zhishuai Zhang, Jiahui Yu, Christian Frank, et~al.
\newblock Noise2music: Text-conditioned music generation with diffusion models.
\newblock \emph{arXiv preprint arXiv:2302.03917}, 2023{\natexlab{b}}.

\bibitem[Kilgour et~al.(2018)Kilgour, Zuluaga, Roblek, and
  Sharifi]{kilgour2018fr}
Kevin Kilgour, Mauricio Zuluaga, Dominik Roblek, and Matthew Sharifi.
\newblock Fr$\backslash$'echet audio distance: A metric for evaluating music
  enhancement algorithms.
\newblock \emph{arXiv preprint arXiv:1812.08466}, 2018.

\bibitem[Koutini et~al.(2021)Koutini, Schl{\"u}ter, Eghbal-zadeh, and
  Widmer]{koutini2021efficient}
Khaled Koutini, Jan Schl{\"u}ter, Hamid Eghbal-zadeh, and Gerhard Widmer.
\newblock Efficient training of audio transformers with patchout.
\newblock \emph{arXiv preprint arXiv:2110.05069}, 2021.

\bibitem[Ribeiro et~al.(2011)Ribeiro, Flor{\^e}ncio, Zhang, and
  Seltzer]{ribeiro2011crowdmos}
Fl{\'a}vio Ribeiro, Dinei Flor{\^e}ncio, Cha Zhang, and Michael Seltzer.
\newblock Crowdmos: An approach for crowdsourcing mean opinion score studies.
\newblock In \emph{2011 IEEE international conference on acoustics, speech and
  signal processing (ICASSP)}, pages 2416--2419. IEEE, 2011.

\bibitem[ITU-R(2017)]{itur2012algorithms}
ITU-R.
\newblock Algorithms to measure audio programme loudness and true-peak audio
  level, 2017.

\bibitem[Lakhotia et~al.(2021)Lakhotia, Kharitonov, Hsu, Adi, Polyak, Bolte,
  Nguyen, Copet, Baevski, Mohamed, et~al.]{lakhotia2021generative}
Kushal Lakhotia, Eugene Kharitonov, Wei-Ning Hsu, Yossi Adi, Adam Polyak,
  Benjamin Bolte, Tu-Anh Nguyen, Jade Copet, Alexei Baevski, Abdelrahman
  Mohamed, et~al.
\newblock On generative spoken language modeling from raw audio.
\newblock \emph{Transactions of the Association for Computational Linguistics},
  9:\penalty0 1336--1354, 2021.

\bibitem[Dong et~al.(2018)Dong, Hsiao, Yang, and Yang]{dong2018musegan}
Hao-Wen Dong, Wen-Yi Hsiao, Li-Chia Yang, and Yi-Hsuan Yang.
\newblock Musegan: Multi-track sequential generative adversarial networks for
  symbolic music generation and accompaniment.
\newblock In \emph{Proceedings of the AAAI Conference on Artificial
  Intelligence}, volume~32, 2018.

\bibitem[Bassan et~al.(2022)Bassan, Adi, and
  Rosenschein]{bassan2022unsupervised}
Shahaf Bassan, Yossi Adi, and Jeffrey~S Rosenschein.
\newblock Unsupervised symbolic music segmentation using ensemble temporal
  prediction errors.
\newblock \emph{arXiv preprint arXiv:2207.00760}, 2022.

\bibitem[Ycart et~al.(2017)Ycart, Benetos, et~al.]{ycart2017study}
Adrien Ycart, Emmanouil Benetos, et~al.
\newblock A study on lstm networks for polyphonic music sequence modelling.
\newblock ISMIR, 2017.

\bibitem[Ji et~al.(2020)Ji, Luo, and Yang]{ji2020comprehensive}
Shulei Ji, Jing Luo, and Xinyu Yang.
\newblock A comprehensive survey on deep music generation: Multi-level
  representations, algorithms, evaluations, and future directions.
\newblock \emph{arXiv preprint arXiv:2011.06801}, 2020.

\bibitem[Dhariwal et~al.(2020)Dhariwal, Jun, Payne, Kim, Radford, and
  Sutskever]{dhariwal2020jukebox}
Prafulla Dhariwal, Heewoo Jun, Christine Payne, Jong~Wook Kim, Alec Radford,
  and Ilya Sutskever.
\newblock Jukebox: A generative model for music.
\newblock \emph{arXiv preprint arXiv:2005.00341}, 2020.

\bibitem[Gan et~al.(2020)Gan, Huang, Chen, Tenenbaum, and
  Torralba]{gan2020foley}
Chuang Gan, Deng Huang, Peihao Chen, Joshua~B Tenenbaum, and Antonio Torralba.
\newblock Foley music: Learning to generate music from videos.
\newblock In \emph{Computer Vision--ECCV 2020: 16th European Conference,
  Glasgow, UK, August 23--28, 2020, Proceedings, Part XI 16}, pages 758--775.
  Springer, 2020.

\bibitem[Huang et~al.(2022)Huang, Jansen, Lee, Ganti, Li, and
  Ellis]{huang2022mulan}
Qingqing Huang, Aren Jansen, Joonseok Lee, Ravi Ganti, Judith~Yue Li, and
  Daniel~PW Ellis.
\newblock Mulan: A joint embedding of music audio and natural language.
\newblock \emph{arXiv preprint arXiv:2208.12415}, 2022.

\bibitem[Maina(2023)]{maina2023msanii}
Kinyugo Maina.
\newblock Msanii: High fidelity music synthesis on a shoestring budget.
\newblock \emph{arXiv preprint arXiv:2301.06468}, 2023.

\bibitem[Rombach et~al.(2022)Rombach, Blattmann, Lorenz, Esser, and
  Ommer]{rombach2022high}
Robin Rombach, Andreas Blattmann, Dominik Lorenz, Patrick Esser, and Bj{\"o}rn
  Ommer.
\newblock High-resolution image synthesis with latent diffusion models.
\newblock In \emph{Proceedings of the IEEE/CVF Conference on Computer Vision
  and Pattern Recognition}, pages 10684--10695, 2022.

\bibitem[Yang et~al.(2022)Yang, Yu, Wang, Wang, Weng, Zou, and
  Yu]{yang2022diffsound}
Dongchao Yang, Jianwei Yu, Helin Wang, Wen Wang, Chao Weng, Yuexian Zou, and
  Dong Yu.
\newblock Diffsound: Discrete diffusion model for text-to-sound generation.
\newblock \emph{arXiv preprint arXiv:2207.09983}, 2022.

\bibitem[Radford et~al.(2021)Radford, Kim, Hallacy, Ramesh, Goh, Agarwal,
  Sastry, Askell, Mishkin, Clark, et~al.]{radford2021learning}
Alec Radford, Jong~Wook Kim, Chris Hallacy, Aditya Ramesh, Gabriel Goh,
  Sandhini Agarwal, Girish Sastry, Amanda Askell, Pamela Mishkin, Jack Clark,
  et~al.
\newblock Learning transferable visual models from natural language
  supervision.
\newblock In \emph{International conference on machine learning}, pages
  8748--8763. PMLR, 2021.

\bibitem[D{\'e}fossez et~al.(2019)D{\'e}fossez, Usunier, Bottou, and
  Bach]{defossez2019music}
Alexandre D{\'e}fossez, Nicolas Usunier, L{\'e}on Bottou, and Francis Bach.
\newblock Music source separation in the waveform domain.
\newblock \emph{arXiv preprint arXiv:1911.13254}, 2019.

\bibitem[Kumar et~al.(2023)Kumar, Seetharaman, Luebs, Kumar, and
  Kumar]{kumar2023high}
Rithesh Kumar, Prem Seetharaman, Alejandro Luebs, Ishaan Kumar, and Kundan
  Kumar.
\newblock High-fidelity audio compression with improved rvqgan.
\newblock \emph{arXiv preprint arXiv:2306.06546}, 2023.

\end{thebibliography}

\clearpage
\appendix
\counterwithin{figure}{section}
\counterwithin{table}{section}

\section{Appendix}
\begin{figure*}[h]
    \centering
    \includegraphics[width=0.9\textwidth]{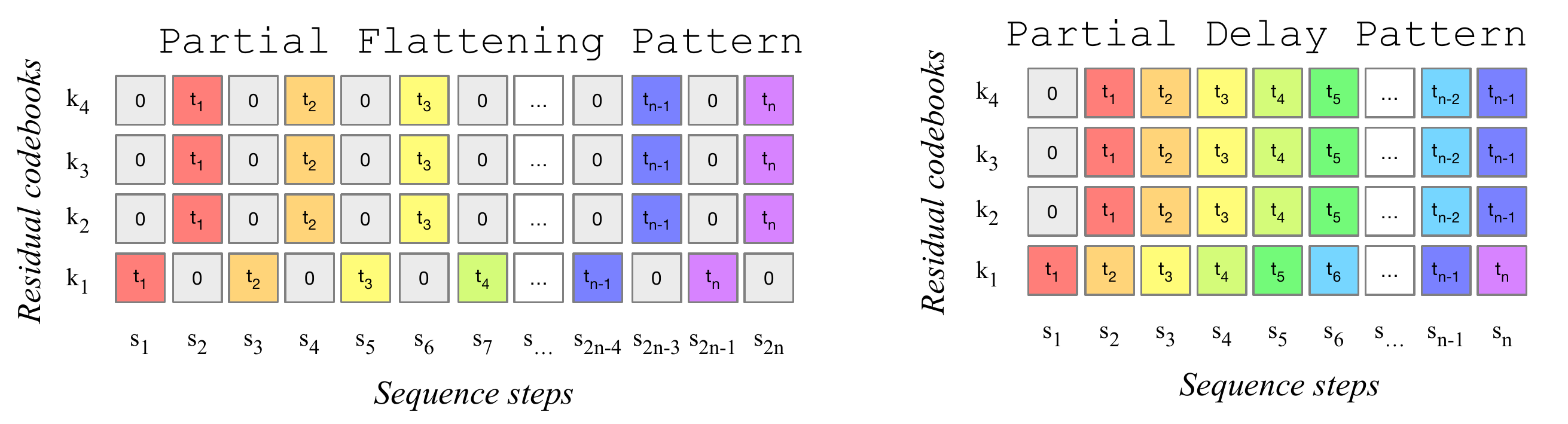}
    \caption{Visualizing partial flattening and partial delays codebook patterns applied on a sequence with 4 parallel streams of quantized values (corresponding to $k_1$, \ldots, $k_4$) and $N$ time steps ($t_1$, \ldots, $k_n$). ``Partial flattening'' separates the first codebook in dedicated steps and interleaves them with the parallel sampling of codebooks 2, 3, and 4, leading the number of interleaved sequences steps $M$ to be twice the number of original steps $N$. The ``partial delay'' pattern consists in delaying by the same amount the codebooks 2, 3, and 4, in our case we use a delay of $1$. The total number of steps of the interleave sequences is $N$ (excluding the last step for simplicity).  }
    \label{fig:pattern_abl}
\end{figure*}

\begin{figure*}[b]
    \centering
    \includegraphics[width=1.0\textwidth]{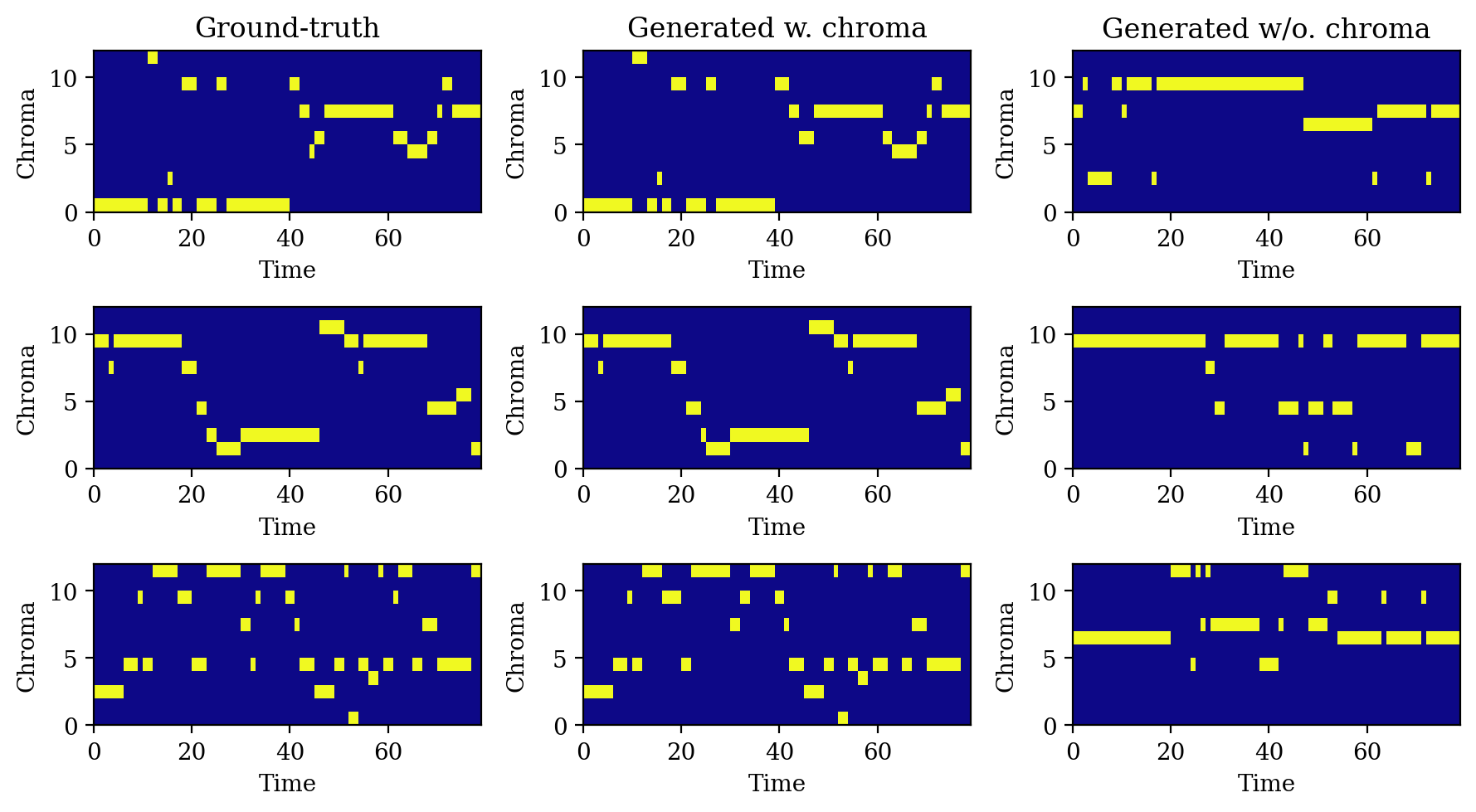}
    \caption{Visualization of quantized chromagram bins over time from reference melody (left), generated music conditioned on chroma and text (middle) and generated music with text-only conditioning (right). Each row is generated using a different chroma condition, all rows share the same text condition: ``90s rock song with electric guitar and heavy drums''. We observe strong adherence to the input melody for the music samples generated with chroma conditioning while rendering novel styles guided by the input text.}
    \label{fig:chromacond}
\end{figure*}

\begin{figure*}[b]
    \centering
    \includegraphics[width=0.5\textwidth]{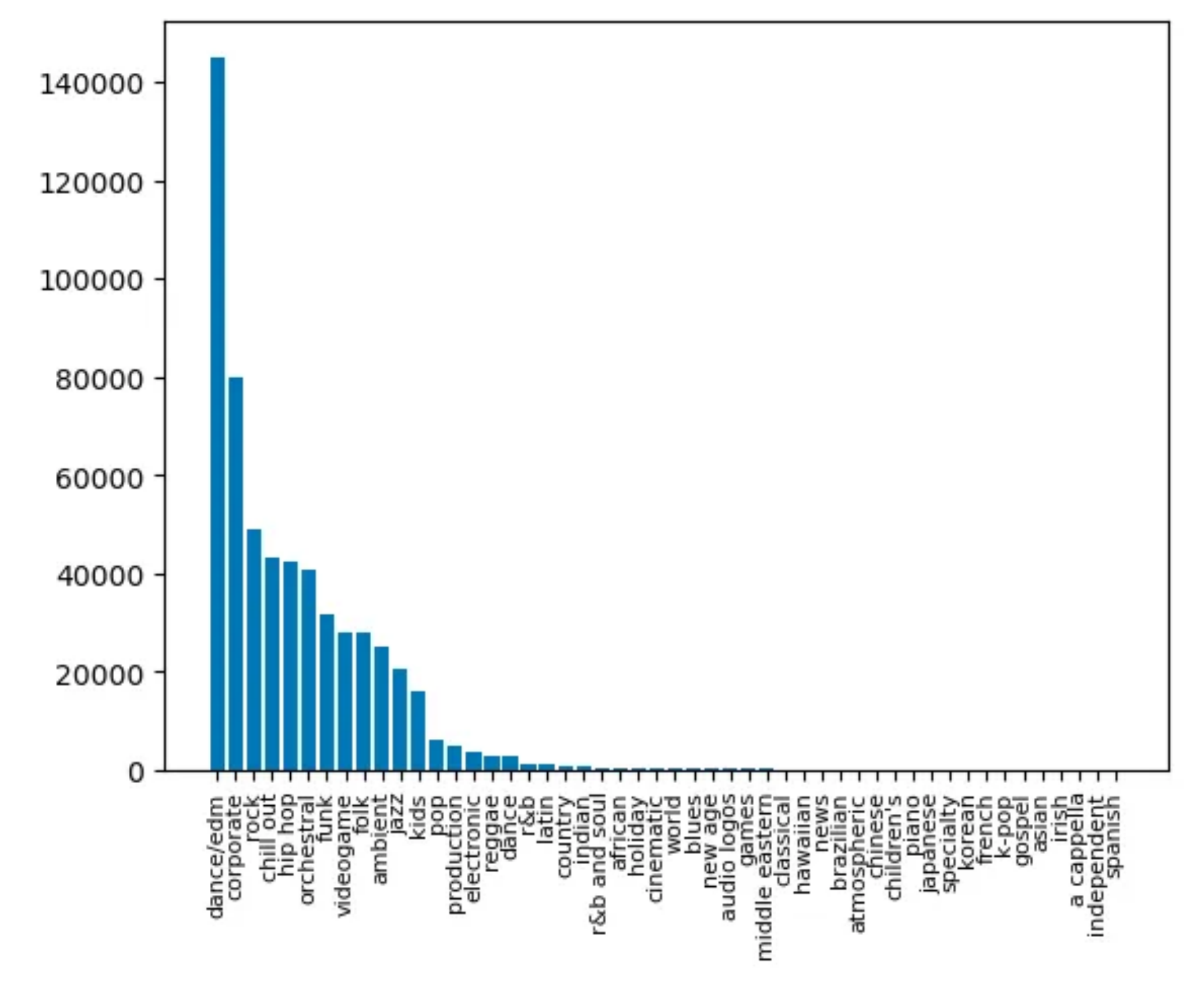}
    \caption{Histogram of the top 50 musical genres in the training dataset.}
    \label{fig:gen}
\end{figure*}

\subsection{Experimental details}
\label{app:sec:experimental}

\newparagraph{Codebook interleaving patterns.} \cref{fig:pattern_abl} provides a visual description of the additional codebook patterns introduced for the ablation in \cref{sec:res}, namely ``partial flattening'' and ``partial delay'' patterns. The intuition behind such patterns is driven by the fact that the first codebook from RVQ is the most important one and predicting the rest of the codebooks in parallel would allow to limit the introduced flattening or delay while maintaining good modeling performance. 

\newparagraph{Melody conditioning.} In this work, we provide an unsupervised approach for melody conditioning through conditioning on the chromagram representation. As shown in \cref{fig:chromacond}, chromagram-based conditioning successfully preserves the melodic structure when generating novel music samples. In preliminary experiments, we noticed that the chromagram tends to be dominated by the lower frequency instruments, mainly by the drums and bass. To mitigate that, we used Demucs \citep{defossez2019music} to first decompose the reference track into four components: drums, bass, vocals, and other. Next, we omit the drums and bass to recover the melodic structure of the residual waveform. Finally, we extract the quantized chromagram to create the conditioning that is later fed to the model.

\newparagraph{Distribution of genres.}
We provide in Figure~\ref{fig:gen} a histogram of the top 50 musical genres present in the dataset.
We notice a clear dominance of the Dance/EDM genre, which in our experience is also one of the genres that is best generated by \method. While we tried to explore a number of resampling scheme to boost the importance of other genres, we observed that oversampling less represented genres would often lead to a worse model overall.

\subsection{Additional experimental results}
\label{app:sec:results}
We provide further ablation studies on the core components of \method, namely the text encoder used for text conditioning described in \cref{sec:conditioning}, text augmentation strategies presented in \cref{sec:hyperparams}, and the used audio tokenization model.  We report results on the MusicCaps dataset to better understand out-of-domain generalization abilities of the different approaches. Finally, we share additional experimental results on optimization methods.

\newparagraph{The effect of text encoder.}
\label{sec:text_enc}
We investigate the impact of the text encoder, comparing three different encoders: T5~\citep{raffel2020t5} \footnote{\url{https://huggingface.co/t5-base}}, Flan-T5~\citep{chung2022scaling}\footnote{\url{https://huggingface.co/google/flan-t5-base}} and CLAP~\citep{laionclap2023} \footnote{\url{https://huggingface.co/lukewys/laion_clap/blob/main/music_audioset_epoch_15_esc_90.14.pt}} with a quantization bottleneck.  
For the CLAP-based encoder, similarly to \citet{agostinelli2023musiclm} we rely on the music embeddings during training and the text embeddings at inference time and we train a RVQ layer on top of the extracted embeddings. More specifically, we first normalize the embeddings and use RVQ with 12 quantizers, each with a codebook size of 1024. Quantizing the CLAP embeddings leads to a homogeneous representation with the discrete tokens further reducing the gap between the audio encoding used at train time and text encoding at test time. We report results for the different text encoders in  \cref{tab:txt_enc}. Both T5 and Flan-T5 perform similarly in terms of objective metrics, the overall quality being slighly better for T5. The CLAP-based model however suffers worse objective and subjective metrics, with the exception of the CLAP score which rely on the same underlying audio and text encoder.

\begin{table}[t!]
  \caption{Text encoder results. We report results for T5, Flan-T5, and CLAP as text encoders. We observe similar results for T5 and Flan-T5 on all the objective metrics. Note that T5 is the text encoder used for the main \method models.
  CLAP encoder performs consistently worse on all the metrics but CLAP score. All comparisons are done with a 300M \method model using text conditioning only.\label{tab:txt_enc}}  
  \centering
  \begin{tabular}{lccc|cc}
    \toprule
             \multicolumn{1}{c}{}& \multicolumn{5}{c}{\footnotesize\textsc{MusicCaps} Test Set}\\ \cmidrule{2-6}
    \textsc{Model} & \textsc{Fad}$_{vgg} \downarrow$      & \textsc{Kl} $\downarrow$ & \textsc{Clap}$_{scr} \uparrow$  & \textsc{Ovl.} $\uparrow$ & \textsc{Rel.} $\uparrow$\\
    \midrule
    T5           & \textbf{3.12}  & \textbf{1.29} & 0.31 & 85.04\pmr{1.23} & \textbf{87.33\pmr{1.9}} \\
    Flan-T5      & 3.36           & 1.30  & 0.32 & \textbf{85.54\pmr{1.01}} & 85.00\pmr{1.63}\\    
    CLAP         & 4.16             & 1.36  &  0.35 & 82.13\pmr{1.29} & 83.56\pmr{1.54} \\
    CLAP (no normalization)         & 4.14             & 1.38  &  0.35 & 84.87\pmr{1.25} & 85.06\pmr{1.72} \\
    CLAP (no quantization)         & 5.07            & 1.37  &  \textbf{0.37} & 84.13\pmr{1.02} & 84.67\pmr{1.42} \\
    \bottomrule
  \end{tabular}
\end{table}

\newparagraph{The effect of text augmentations.}
\label{sec:text_aug}
We examine the impact of text augmentation strategies for the proposed method. In particular, we study the use of condition merging (i.e. concatenating additional metadata to the text description), text normalization (text-norm.) and word dropout. We report objective metrics for the different augmentation strategies in \cref{tab:txt_strat}. We observe a gain in FAD and KL when leveraging the additional metadata with condition merging. However, neither text normalization or word dropout improves the results in terms of objective metrics. 

\begin{table}[t!]
  \caption{Text augmentations strategies results. We report objective metrics using only the original text description (no augmentation) and for different text augmentation strategies: using condition merging to augment the text description with metadata, using text-normalization (text-norm.) and applying word dropout on the resulting text. We use 300M \method models trained for 500K steps. Condition merging improves the result over training only over the original text description. Other augmentations perform worst on all metrics. We use the Condition Merging with Word dropout, showing the best text relevance, in our main models. \label{tab:txt_strat}}  
  \centering
\resizebox{0.9\columnwidth}{!}{
  \begin{tabular}{lccc|cc}
    \toprule
    \multicolumn{1}{c}{} & \multicolumn{5}{c}{\footnotesize \textsc{MusicCaps} Test Set}\\ 
     \cmidrule{2-6}
    \textsc{Configuration} & \textsc{Fad}$_{vgg} \downarrow$      & \textsc{Kl} $\downarrow$ & \textsc{Clap}$_{scr}$ &
    \textsc{Ovl.} $\uparrow$ & \textsc{Rel.} $\uparrow$ \\
    \midrule
    No augmentation                    & 3.68  & 1.28 & \textbf{0.31} & \textbf{83.40}\pmr{1.44} & 81.16\pmr{1.29}\\
    Condition Merging (CM)             & \textbf{3.28}  & \textbf{1.26} & \textbf{0.31} & 82.60\pmr{1.41} & 84.45 \pmr{1.16}\\ 
    CM + Text-norm. (TN)               & 3.78  & 1.30 & 0.29 & 80.57\pmr{2.14} & 82.40\pmr{1.09}\\
    CM+ Word dropout (WD)              & 3.31  & 1.31 & 0.30 & 82.52\pmr{1.55} & \textbf{85.27}\pmr{0.97}\\
    CM + TN + WD  & 3.41  & 1.39 & 0.30 & 81.18\pmr{1.91} & 84.32 \pmr{1.59}\\
    \bottomrule
  \end{tabular}}
\end{table}

\newparagraph{The effect of the audio tokenizer.}
\label{sec:audio_tokenizer}
\looseness=-1
We experiment with replacing EnCodec with Descript Audio Codec (DAC)~\citep{kumar2023high}\footnote{Using the public implementation \href{https://github.com/descriptinc/descript-audio-codec/}{github.com/descriptinc/descript-audio-codec}.}, a similar audio compression models, that uses a different training
set, a similar adversarial loss enhanced with multiband discriminators, but performs quantization in a lower dimension space to
improve codebook usage. DAC compresses audio at 44.1 kHz with 9 codebooks, and a framerate of 86 Hz.
We trained a small (300M parameters) and a medium (1.5B parameters) \method model using both DAC and EnCodec as an audio tokenizer,
on a vocal-free version of our dataset. 
The results provided in Table~\ref{tab:dac} show a worse FAD and KL on our in domain test set. On MusicCaps, the FAD is improved by using DAC, although the KL is worse, as well as the subjective evaluations. The EnCodec model used in this work was specifically designed to operate at a lower frame rate (50 Hz) than DAC (86 Hz), thus reducing by 40\% the inference runtime for producing the same audio. Note finally that DAC was trained on a different dataset than EnCodec, and further experiments would be required to understand exactly what influences the fitness of such compression models to be used with auto-regressive language models.

\begin{table}[t!]
    \centering
      \caption{We test replacing EnCodec with DAC~\citep{kumar2023high} using their implementation.
      DAC is a 44.1 kHz model with 9 codebooks and a frame rate of 86 Hz.
      Those models are trained on a vocal-free version of our dataset, hence
      the objective metrics will not match those reported in the other tables.
      We report objective metrics both on our in domain test set and MusicCaps~\citep{agostinelli2023musiclm},
      and subjective metrics only on MusicCaps.
        \label{tab:dac}}  
      {\centering
      \begin{tabular}{l|cc|cc|c}
        \toprule
           \multicolumn{1}{c}{}& \multicolumn{2}{c}{\footnotesize In Domain Test Set}&
           \multicolumn{3}{c}{\footnotesize\textsc{MusicCaps} Test Set}
           \\\cmidrule{2-6}
            \textsc{Model}  & \textsc{Fad}$_{\text{vgg}} \downarrow$      & \textsc{Kl} $\downarrow$   & 
            \textsc{Fad}$_{\text{vgg}} \downarrow$      & \textsc{Kl} $\downarrow$   & \textsc{Ovl.} $\uparrow$ \\
        \midrule
        \method+ DAC small  & 3.45 & 0.58 & 4.46 & 1.35 & 83.32\pmr{0.95}  \\
        \method+ DAC medium & 2.42 & 0.57 & \textbf{4.32} & 1.30 & 84.46\pmr{0.97}  \\
        \midrule
        \method+ EnCodec small & 0.67 & 0.54 & 5.26 & 1.27 & 84.69\pmr{0.90}  \\
        \method+ EnCodec medium & \textbf{0.49} & \textbf{0.52} & 5.05 & \textbf{1.23} & \textbf{86.09}\pmr{0.88} \\
        \bottomrule
      \end{tabular}}
\end{table}

\newparagraph{Effect of D-Adaptation.}

D-Adaptation is a novel automated way of selecting the overall learning rate of the Adam optimizer, i.e. its $\alpha$ parameter, dynamically throughout the training, introduced by~\citet{defazio2023learning}. We observed improved convergence for the 300M parameter model, although for larger models, e.g. 1.5B and 3.3B, we observed the automated rule led to deteriorated performance, both on the train and validation set. Further investigation is required to better understand the effect of D-Adaptation, and whether it can scale to the largest model. The convergence
for both methods can be observed on the train and validation set in Figure~\ref{fig:dadam_vs_adam}.

\begin{figure}
\centering
\begin{subfigure}[t]{.4\textwidth}
\centering
  \begin{tikzpicture}[
    adam/.style={color=brown, solid,mark options={fill=yellow}},
    dadam/.style={color=teal, solid,mark options={fill=green}},
    large/.style={mark=*},
    gpt/.style={mark=triangle*},
    every node/.style={font=\fontsize{8}{5}\selectfont}
  ]
    \begin{axis}[
        xlabel={Epoch},
        ylabel={Train Perplexity},
        name=lrplot,
        yticklabel pos=left,
        style={font=\footnotesize},
        width=\linewidth,
        ]

        \pgfplotstableread{figs/0accc250.tsv}\tablea
        \pgfplotstableread{figs/0ae1a06f.tsv}\tableb
        \pgfplotstableread{figs/8b8fd412.tsv}\tablec
        \pgfplotstableread{figs/c127f9c3.tsv}\tabled
        \addplot+[dadam,large] table[x=epoch, y=tppl, each nth point=20, filter discard warning=false, unbounded coords=discard] {\tablea};
        \addplot+[adam,large] table[x=epoch, y=tppl, each nth point=20, filter discard warning=false, unbounded coords=discard] {\tablec};
        \addplot+[dadam,gpt] table[x=epoch, y=tppl, each nth point=20, filter discard warning=false, unbounded coords=discard] {\tableb};
        \addplot+[adam,gpt] table[x=epoch, y=tppl, each nth point=20, filter discard warning=false, unbounded coords=discard] {\tabled};
    \end{axis}
\end{tikzpicture}
\end{subfigure} \hspace{2.2cm}
\begin{subfigure}[t]{.4\textwidth}
\centering
  \begin{tikzpicture}[
    adam/.style={color=brown, solid,mark options={fill=yellow}},
    dadam/.style={color=teal, solid,mark options={fill=green}},
    large/.style={mark=*},
    gpt/.style={mark=triangle*},
    every node/.style={font=\fontsize{8}{5}\selectfont}
  ]
    \begin{axis}[
        xlabel={Epoch},
        ylabel={Valid Perplexity},
        name=lrplot,
        yticklabel pos=right,
        ylabel near ticks,
        style={font=\footnotesize},
        width=\linewidth,
        legend entries={
            {{DAdam, 300M},
             {Adam, 300M},
             {DAdam, 1.5B},
             {Adam, 1.5B},
             }
        },
        legend style={
            overlay,
            font=\fontsize{5}{5}\selectfont,at={(-0.05,0.5)},anchor=east,
            legend columns=1, fill=white,draw=black}, 
        ]

        \pgfplotstableread{figs/0accc250.tsv}\tablea
        \pgfplotstableread{figs/0ae1a06f.tsv}\tableb
        \pgfplotstableread{figs/8b8fd412.tsv}\tablec
        \pgfplotstableread{figs/c127f9c3.tsv}\tabled
        \addplot+[dadam,large] table[x=epoch, y=vppl, each nth point=20, filter discard warning=false, unbounded coords=discard] {\tablea};
        \addplot+[adam,large] table[x=epoch, y=vppl, each nth point=20, filter discard warning=false, unbounded coords=discard] {\tablec};
        \addplot+[dadam,gpt] table[x=epoch, y=vppl, each nth point=20, filter discard warning=false, unbounded coords=discard] {\tableb};
        \addplot+[adam,gpt] table[x=epoch, y=vppl, each nth point=20, filter discard warning=false, unbounded coords=discard] {\tabled};
    \end{axis}
\end{tikzpicture}
\end{subfigure}
\caption{Comparison of Adam and Adam with D-Adaptation~\citep{defazio2023learning}. While D-Adaptation provided
consistent gains for the 300M parameters model, we observed worse convergence both on the train (left) and validation (right) set
for the 1.5B parameters model.
}
\label{fig:dadam_vs_adam}
\end{figure}

\end{document}